\documentclass[useAMS,usenatbib]{mn2e}  
\usepackage{times}

\def\chandra{{\it Chandra\/}}

\def\heao1{{\it HEAO-1\/}}
\def\hst{{\it {\it HST}\/}}
\def\iras{{\it IRAS\/}}
\def\spitzer{{\it Spitzer\/}}

 \def\herschel{{\it Herschel\/}}
\def\irs{{\it IRS\/}}

\def\swift{{\it Swift\/}}

\def\herschel{{\it Herschel\/}}
\def\iras{{\it IRAS\/}}

\newcommand{\ltsima}{$\; \buildrel < \over \sim \;$}
\newcommand{\simlt}{\lower.5ex\hbox{\ltsima}}
\newcommand{\gtsima}{$\; \buildrel > \over \sim \;$}
\newcommand{\simgt}{\lower.5ex\hbox{\gtsima}}

\newcommand{\cgs}{ ${\rm erg~cm}^{-2}~{\rm s}^{-1}$} 
\newcommand{\lum}{\rm erg~s$^{-1}$}

\def\lesssim{\mathrel{\hbox{\rlap{\hbox{\lower4pt\hbox{$\sim$}}}\hbox{$<$}}}}
\def\gtrsim{\mathrel{\hbox{\rlap{\hbox{\lower4pt\hbox{$\sim$}}}\hbox{$>$}}}}

\def\xray{\hbox{X-ray}}

\def\micron{\hbox{$\mu$m}}

\usepackage{graphicx}
\usepackage{txfonts}
\usepackage{rotating}

\title[The AGN content in luminous IR galaxies at $z{\sim}$2]
{The AGN content in luminous IR galaxies at $z{\sim}$2 from a global SED 
analysis including Herschel data}

\author[F. Pozzi et al.]{F. Pozzi,$^{1,2}$\thanks{E-mail:f.pozzi@unibo.it} 
C. Vignali,$^{1,2}$ C. Gruppioni,$^{2}$ A. Feltre,$^{3}$ 
J.Fritz,$^{4}$ D. Fadda,$^{5}$ P. Andreani,$^{6}$ \
\newauthor S. Berta,$^{7}$ A. Cimatti,$^{1}$ I. Delvecchio,$^{1,2}$ 
D. Lutz,$^{7}$ B. Magnelli,$^{7}$ R. Maiolino,$^{8}$ 
R. Nordon,$^{7}$ \newauthor P. Popesso,$^{7}$\ 
G. Rodighiero,$^{3}$ D. Rosario,$^{7}$ P. Santini,$^{7,8}$ 
M. Vaccari$^{3}$\\
$^{1}$Dipartimento di Astronomia, Universit\`a degli Studi di Bologna, 
Via Ranzani 1, I--40127 Bologna, Italy\\
$^{2}$INAF --- Osservatorio Astronomico di Bologna, Via Ranzani 1, I--40127 
Bologna, Italy\\
$^{3}$Dipartimento di Astronomia, Vicolo Osservatorio 3, 35122 Padova,
Italy\\
$^{4}$Sterrenkundig Observatorium, Universiteit Gent, Krijgslaan 281 S9,
B-9000 Gent, Belgium\\
$^{5}$IPAC, California Institute of Technology, Pasadena, CA 91125, USA\\
$^{6}$European Southern Observatory, Karl-Schwarzschild-Str. 2, 85748 
Garching, Germany\\
$^{7}$Max-Plank-Institut f\"ur extraterrestrische Physik, Postfach 1312,
85741 Garching, Germany\\
$^{8}$INAF --- Osservatorio Astronomico di Roma, via di Frascati 33,
00040 Monte Porzio Catone, Italy}

\begin{document}

\date{Accepted ... Received ...; in original form ...}


\maketitle

\label{firstpage}

\begin{abstract}

We use \herschel-PACS far-infrared data, combined with previous multi-band 
information and mid-IR spectra, to properly account for the presence of an 
active nucleus and constrain its energetic contribution in luminous 
infrared (IR) sources at $z\sim2$. The sample is composed of 
24 sources in the GOODS-South field, with typical IR luminosity of 
$10^{12}~L_{\odot}$. Data from the 4~Ms \chandra\ X-ray imaging in this field 
are also used to identify and characterize AGN emission. \\
We reproduce the observed spectral energy distribution (SED), decomposed into 
a host-galaxy and an AGN component. A smooth-torus model for
circum-nuclear dust is used to account for the direct and re-processed 
contribution from the AGN. \\
We confirm that galaxies with typical $L_{8-1000{\mu}m}{\sim}10^{12}L_{\odot}$ at 
$z{\sim}2$ are powered predominantly by star-formation. An AGN 
component is present in nine objects (${\sim}35\%$ of the sample) at 
the 3${\sigma}$ confidence level, but its contribution to the 
8-1000~${\mu}$m emission accounts for only ${\sim}5\%$ of the energy budget. 
The AGN contribution rises to $\sim$23\% over the 5--30~${\mu}$m range 
(in agreement with \spitzer\ IRS results) and to $\sim$60\% over the narrow 
2--6~${\mu}$m range. 
The presence of an AGN is confirmed by \xray\ data for 3 (out of nine) 
sources, with \xray\ spectral analysis indicating the presence of 
significant absorption, i.e. N$_{H}{\sim}10^{23}-10^{24}$~cm$^{-2}$. 
An additional source shows indications of obscured AGN emission from 
\xray\ data.
The comparison between the mid-IR--derived \xray\ luminosities and those 
obtained from \xray\ data suggests that obscuration is likely present also 
in the remaining six sources that harbour an AGN according to the 
SED-fitting analysis.
\end{abstract}
 

%

\section{Introduction}
\label{intro_sec}
Ultra-luminous infrared galaxies (ULIRGs, defined as having 
$L_{IR}>10^{12}L_{\odot}$; \citealt{1987ARA&A..25..187S}) are among the most 
luminous objects of the Universe, radiating most of their energy in the 
infrared (IR) band. 
In the local Universe, ULIRGs are rare objects and, although very luminous, 
account for only ${\sim}5\%$ of the total integrated IR luminosity density 
\citep{1991AJ....101..354S}. Moving to higher redshift, the contribution of 
ULIRGs to the energy budget increases, as clearly illustrated by infrared 
luminosity function studies (\citealt{2005ApJ...632..169L}; 
\citealt{2007ApJ...660...97C}). 
\citet{2011A&A...528A..35M}, deriving the infrared luminosity function 
up to $z{\sim}2.3$ from deep 24 and 70~${\mu}$m \spitzer\ data, obtained 
a contribution of ${\sim}17 \%$ from ULIRGs to the IR luminosity density 
at $z{\sim}2$. This result has been recently confirmed also by the 
\herschel\ Science Demonstration Phase (SDP) preliminary results; 
\cite{2010A&A...518L..27G}, by deriving the total IR luminosity function up 
to $z\sim{3}$, estimated that ULIRGs account for ${\sim}30 \%$ to the IR 
luminosity density at $z{\sim}2$.

At high redshift ($z{\sim}2$), a key question regards the nature of the 
sources that power these ultra-luminous objects (i.e., star-formation 
vs. accretion). Locally, thanks to {\it Infrared Space observatory 
(ISO)} mid-IR spectroscopy, ULIRGs were proven to be powered predominantly 
by star formation in the mid-IR, 
with the fraction of AGN-powered objects increasing with 
luminosity (from ${\sim}$15\% at $L_{IR}<2{\times}10^{12}~L_{\odot}$ to about 
50\% at higher luminosity; \citealt{1998ApJ...505L.103L}). A limited 
percentage (15--20\%) of mid-IR light dominated by accretion processes
has also been found using $L$-band (3-4~$\mu$m) observations of local ULIRGs 
with $L_{IR}\sim10^{12}L_{\odot}$ by \cite{2010MNRAS.401..197R}. 
Interestingly, these authors show that, although their sources are powered 
mostly by starburst processes, at least 60\% of them contain 
an active nucleus. 
Using accurate optical spectral line diagnostics applied to a sample
of 70~$\mu$m selected luminous sources at $z\sim$1, \cite{2010MNRAS.403.1474S} 
found that only 20--30\% of their objects may host an active nucleus. 
However, such AGN are never bolometrically dominant. 
In the same paper, a discussion on how the low AGN incidence can be partially 
due to selection effects (i.e. the 70~$\mu$m band sampling the starburst 
50~$\mu$m rest-frame far-IR bump) is also presented. 

At $z\sim2$, our understanding of the AGN content in ULIRGs is much more 
uncertain. On the one hand, 
sources selected with a 24~${\mu}$m flux density $\simgt$0.9--1~mJy mainly 
sample the bright end of the ULIRG population. The analysis of the IRS 
spectra of a color-selected sample of sources with 
S$_{24{\mu}m}{\simgt}0.9$~mJy and at $z>1.5$ 
($L_{IR}{\sim}7{\times}10^{12}L_{\odot}$) shows that the majority of these 
sources are AGN-dominated (${\sim}75\%$; \citealt{2008ApJ...683..659S}). 
On the other hand, sub-millimeter selected galaxies (SMGs) at $z{\sim}2$, 
falling in the ULIRG regime, appear largely starburst-dominated objects 
(e.g., \citealt{2007ApJ...660.1060V}; \citealt{2008ApJ...675.1171P}).

In this paper, we aim at providing a better understanding of luminous
sources at $z\sim$2 with typical IR luminosities of $10^{12}L_{\odot}$, 
i.e. sampling the knee of the IR luminosity function instead of the bright end. 
We will re-analyse the sample from \citet[hereafter F10]{2010ApJ...719..425F}, 
where an estimate of the AGN contribution in ULIRGs has already been presented, 
based on \spitzer\ IRS data, \chandra\ 2~Ms observations, 
optical/near-IR multi-band properties, and ACS morphological properties. 
In this work, we will extend the analysis to far-IR data, obtained recently 
by the \herschel\ satellite as part of the guaranteed survey 
``PACS Evolutionary Probe'' (PEP; \citealt{2011A&A...532A..90L}), and
to the recently published \chandra\ 4~Ms data 
(\citealt{2011ApJS..195...10X}).

We note that an analysis of the F10 sample (including the new far-IR data) 
has been already presented by \cite{2012ApJ...745..182N} [hereafter N12], 
where the general properties of the mid-to-far IR spectral energy 
distributions (SEDs) of 0.7$<z<$2.5 galaxies have been
investigated. 
The authors confirm the early \herschel\ results (e.g., 
\citealt{2010A&A...518L..29E}), i.e. the star-formation rates at $z\sim2$ 
are over-estimated if derived from 24~${\mu}$m flux densities, and discuss how 
this effect can be due to enhanced PAH emission with respect to local templates 
(see also \citealt{2010A&A...518L..29E}; \citealt{2011A&A...533A.119E}). In N12, the SED library
from \cite{2001ApJ...556..562C} was used, longward of 6~$\mu$m
rest-frame, and the IRS spectra data were stacked to improve the faint
signal of sources at $z\sim{2}$. 
In this paper we adopt a different method to study the properties of 
$z\sim2$ IR galaxies, as described in $\S$\ref{modeling_sec};  
a comparison of our results vs. those obtained by N12 is presented in 
$\S$\ref{agnfraction_sec_fromir}. 
  
Hereafter, we adopt the concordance cosmology 
($H_{0}=70$~km~sec$^{-1}$~Mpc$^{-1}$, $\Omega_{m}$=0.3, and 
$\Omega_{\Lambda}$=0.7, \citealt{2003ApJS..148..175S}). 
Magnitudes are expressed in the Vega system.

\section{The data}
\label{sample_sec}

The sample is composed of 24 luminous sources at $z{\sim}$2, selected by F10 in 
the \chandra\ Deep Field South (CDF-S), with 24~${\mu}$m flux 
densities $S(24~{\mu}m){\sim}0.14-0.55$ mJy and at $z=1.75-2.40$. 
The sample can be considered luminosity-selected, since all of the 
sources satisfying these selection criteria are considered. 
Given the redshift of the sources, their 24~$\mu$m flux densities translate 
roughly into infrared luminosity around $10^{12}L_{\odot}$. 
We excluded two objects, originally defined as luminous IR galaxies 
(LIRGs; $L_{IR}>10^{11}L_{\odot}$) by F10 (L5511 and L6211) and subsequently 
included in the ULIRG class thanks to the new IRS redshift measurements, since 
they would be the only two sources outside the \hst\ ACS area 
(L5511 is also outside the \herschel\ area).

The sample benefits from a large amount of information 
available for the CDF-S, from multi-band photometry to the recent 
\spitzer\ IRS spectroscopy. The IRS observations were performed
at low-resolution in the observed 14-35~\micron\ wavelength
regime, i.e. sampling at $z\sim2$ the important rest-frame PAH 
features at 6.2~\micron\ and 7.7~\micron, and the 9.7~\micron\ silicate 
feature (see F10). 
As reported above, in this work we also use recent far-IR data obtained 
from the \herschel\ satellite, which consists of 
PACS data at 70, 100 and 160~${\mu}$m from the PEP survey \citep{2011A&A...532A..90L}. 
We use the PACS blind catalogue v1.3 down to 3$\sigma$ limits of 1.2, 1.2 and 
2.0~mJy at 70, 110 and 160 ${\mu}$m, respectively. 
\herschel\ data have been matched to 24~\micron\ \spitzer\ MIPS sources 
\citep{2009A&A...496...57M} using the likelihood ratio technique 
(\citealt{1992MNRAS.259..413S}; \citealt{2001Ap&SS.276..957C}), 
as described in \cite{2011A&A...532A..49B}. Shorter wavelength information has 
been included by matching the 24~\micron\ sources to the multi-band (from 
UV to \spitzer\ IRAC bands) GOODS-MUSIC photometric catalog 
\citep{2009yCat..35040751S}.
In particular, we cross-correlated the 24~\micron\ selected ULIRGs with the PEP 
catalogue, considering the positions of the 24~\micron\ sources already 
associated to the PEP ones. Among the 24 ULIRGs, 21 sources have counterparts 
in PACS. For the 3 sources undetected by \herschel\ (U5050, U5152, and U5153), 
we consider the conservative 5$\sigma$ upper limits (2.2, 2.0 and 3.0 mJy at 
70, 100 and 160~\micron, respectively). For PACS detections, the typical 
separation between 24~\micron\ and PEP sources is less than 1\arcsec, 
with the exception of U5805 and U16526 (${\sim}$4\arcsec). By visually 
checking the far-IR images, these sources appear as blends of more 
than one 24~${\mu}$m source. In these cases, the PACS flux densities are 
considered as upper limits. Photometry at 16~\micron, obtained by F10 
from IRS data, has also been included. 

Finally, we consider the recently published \chandra\ 4~Ms point-like source 
catalogue in the CDF-S (\citealt{2011ApJS..195...10X}). Eight of the 24 sources 
have an \xray\ counterpart within 1\arcsec. 

Table~\ref{sample_prop} lists the source names, redshifts from optical and 
IRS spectroscopy (see F10), source IDs in the GOODS-MUSIC catalogue, 
flux densities and associated errors from mid-IR (16~\micron) to far-IR 
(160~${\mu}$m). 
%

\begin{table*}
\begin{center}
\caption{The sample: multi-band information}
\begin{tabular}{lcccccccc}\hline\hline
  Name    &       $z_{opt}$ &    $z_{IRS}$ &ID$_{MUSIC}$
  &S$_{16{\mu}m}$ & S$_{24{\mu}m}$  &S$_{70{\mu}m}$ &S$_{100{\mu}m}$ &S$_{160{\mu}m}$  \\ 
(1) & (2) & (3) & (4) & (5) & (6) & (7) & (8) & (9) \\ \hline
  U428  &  $-$1.664  &  1.783  &     8053  &     65   $\pm$     3  &     289   $\pm$     5 &   1270  $\pm$ 380   &      2870    $\pm$  450  &      $<3300$      \\
  U4367  &  $-$1.762  &  1.624   &   3723  &    101   $\pm$  8  &     152    $\pm$     4  &           $<1800$                 &    $<1900$                             &     2270    $\pm$    530  \\
  U4451  &  $-$1.684  &  1.875   &   3689  &                &      199        $\pm$     4  &            $<1800$                  &      2000   $\pm$  380 &     5110     $\pm$      550  \\
  U4499  &  $-$1.909  &  1.956   &  70361  &    $<35$         &     167  $\pm$      3 &        $<1800$               &       $<1900$                         &     6990    $\pm$       580  \\
  U4631  &   1.896  &  1.841   &   7087  &    $<21$     &     275     $\pm$    5  &         $<1800$            &          $<1900$                     &     4010  $\pm$      560  \\
  U4639  &   2.130  &  2.112   &   7553  &    $<14$      &     213     $\pm$    4  &        $<1800$                &          $<1900$                        &     3570     $\pm$      550  \\
  U4642  &  $-$1.748  &  1.898  &   8217  &     65    $\pm$      9  &     242    $\pm$     5  &        $<1800$                  &      1150  $\pm$    360  &      $<3300$       \\
  U4812  &   1.910  &  1.930  &  13175  &    $<53$       &     295     $\pm$    6  &   2590  $\pm$  410  &     10000  $\pm$   380  &    24830    $\pm$     730  \\
  U4950  &   2.291  &  2.312   &  15260  &    453     $\pm$     8  &     557    $\pm$     6  &          $<1800$                 &      2060  $\pm$   420  &          $<3300$       \\
  U4958  &   2.145  &  2.118   &  15483  &    109     $\pm$     10  &     232    $\pm$     4  &     1360$\pm$  410  &      2690    $\pm$  380  &     3720    $\pm$    550  \\
  U5050  &  $-$1.720  &  1.938   &  13758  &     51      $\pm$    7  &     197     $\pm$    4  &       $<1800$                      &         $<1900$                           &  $<3300$ \\
  U5059  &  $-$1.543  &  1.769  &  13887  &    $<37$  &     272    $\pm$     5  &     1450  $\pm$  380 &      1750 $\pm$    360  &     4400   $\pm$     610  \\
  U5150  &  $-$1.738  &  1.898  &  15083  &     65    $\pm$     11  &     277   $\pm$      4  &        $<1800$                &     4350    $\pm$  440  &     5690    $\pm$        700  \\
  U5152  &  $-$1.888  &  1.794    &  70066  &     66       $\pm$   9  &     267     $\pm$    8  &    $<1800$                          &            $<1900$                       &   $<3300$ \\
  U5153  &  $-$2.030  &  2.442   &  70054  &    $<34$     &     166    $\pm$     3  &     $<1800$                      &             $<1900$                       &   $<3300$ \\
  U5632  &   1.998  &  2.016  &   9261  &     $83$      $\pm$    2  &     427    $\pm$    5  &     2290  $\pm$  530   &    4250    $\pm$  360  &    11470   $\pm$        560  \\
  U5652  &   1.616  &  1.618   &   6758  &    154      $\pm$    9  &     343    $\pm$    4  &     1600 $\pm$  380   &  6920    $\pm$   380    &    17220   $\pm$        740  \\
  U5775  &  $-$1.779  &  1.897   &   9361  &    $<31$      &     164   $\pm$      4  &   $<1800$                          &      1440  $\pm$    360  &       3930     $\pm$      670  \\
  U5795  &  $-$1.524  &  1.703   &  11201  &    $<31$     &     250   $\pm$     4  &   $<1800$                          &        $<1900$                           &     5030  $\pm$    670  \\
  U5801  &  $-$1.642  &  1.841  &  12003  &    $<35$     &     186    $\pm$     4  &    $<1800$                      &          $<1900$                         &     2310   $\pm$         550  \\
  U5805  &  $-$2.093  &  2.073   &  12229  &     68     $\pm$     6  &     172     $\pm$    4  &   $<1800$                         &    1910$^{a}$    $\pm$    360  &     5100$^{a}$    $\pm$       670  \\
  U5829  &  $-$1.597  &  1.742   &   9338  &     57    $\pm$    5 &     185     $\pm$    4  &      $<1800$                     &      1920   $\pm$   380  &     4320  $\pm$      550  \\
  U5877  &  $-$1.708  &  1.886   &  13250  &    179     $\pm$    8  &     364    $\pm$     5  &     4600     $\pm$   420  &      7460    $\pm$    440  &    11570   $\pm$       870  \\
 U16526  &  $-$1.718  &  1.749   &   3420  &     69     $\pm$     4  &     306     $\pm$    9  &     4130$^{a}$      $\pm$    920  &      8600$^{a}$   $\pm$     390  &    15970 $^{a}$    $\pm$     760  \\\hline
\end{tabular}
\begin{minipage}[h]{13.8cm}
\footnotesize
Notes --- (1) Source name (F10); (2) optical redshift (negative and
positive values are referred to photometric and spectroscopic redshifts, 
respectively; see F10 for details); (3) redshift derived from IRS spectroscopy 
(F10); 
(4) ID from the v2 GOODS-MUSIC catalogue (\citealt{2009yCat..35040751S}); 
(5) 16~\micron\ flux density from F10, in $\mu$Jy; 
(6) 24~\micron\ flux density in $\mu$Jy (\citealt{2009A&A...496...57M}); 
(7), (8), (9) \herschel\ PEP flux density in $\mu$Jy 
(\citealt{2011A&A...532A..90L}). 
$^{a}$ indicates a possible contribution from a nearby source. 
\end{minipage}
\label{sample_prop}
\end{center}
\end{table*}

\section{SED decomposition}
\label{modeling_sec}

The IR energy budget of a galaxy can be mainly ascribed to 
stellar photosphere and star-formation emissions and 
accretion processes; to estimate the relative importance of these three 
processes, a proper SED decomposition should be carried out. 
Disentangling the different contributions to the total SED is becoming 
more and more effective with the advent of the \spitzer\ and \herschel\ 
satellites. In this regard, many studies have been performed to compare 
the full range of observed photometric data with expectations from a 
host-galaxy component and models for the circum-nuclear dust emission (e.g., 
\citealt{2008ApJ...675..960P}; \citealt{2008MNRAS.386.1252H}; 
\citealt{2009MNRAS.395.2189V}; \citealt{2010A&A...517A..11P}).
Here, together with the full-band photometric SED, we benefit also from the IRS 
spectra, that we combine with the photometric datapoints 
(Sec.~\ref{fitting_sec}; see also \citealt{2011MNRAS.414.1082M}, 
\citealt{2011ApJ...736...82A} for examples of combined 
photometric/spectroscopic data analysis). The IRS spectra provide an 
important diagnostic, sampling the rest-frame mid-IR spectral range 
(${\sim}5-12~\mu$m for our sample), where the difference between starburst and 
AGN is strongest. In this wavelength range, starburst galaxies are 
generally characterized by prominent polycyclic aromatic (PAH) features 
and weak 10~\micron\ continuum, whereas AGN display weak or no PAH features 
plus a strong continuum (e.g. \citealt{2000A&A...359..887L}).

\subsection{AGN and stellar components}
\label{torus_comp_sec}
We have decomposed the observed SEDs using three distinct components: 
stars, having the bulk of the emission in the optical/near-IR; hot dust, 
mainly heated by UV/optical emission due to gas accreting onto the 
super-massive black hole (SMBH) and whose emission peaks between 
a few and a few tens of microns; cold dust, principally heated by star 
formation (we refer to \citealt{2010A&A...517A..11P}; but see also 
\citealt{2008MNRAS.386.1252H} for a detailed description 
of the properties of the AGN and host-galaxy (stars$+$cold dust) components). 
Here we report only the most important issues concerning this analysis, 
with the AGN component being the main focus. 

The stellar component has been modelled as the sum of simple stellar 
population (SSP) models of solar metallicity and ages ranging 
from ≈1~Myr to 2.3~Gyr, which corresponds to the time elapsed 
between $z=6$ (the redshift assumed for the initial star formation stars 
to form) and $z\sim{2}$ in the adopted cosmology. A \cite{1955VA......1..283S} 
initial mass function (IMF), with mass in the range 0.15--120 
M$_{\odot}$, is assumed. The SSP spectra have been weighted by a Schmidt-like 
law of star formation (see \citealt{2004A&A...418..913B}): 

\begin {equation}
SFR(t)=\frac{T_{G}-t}{T_{G}}{\times}\exp\left({-\frac{T_{G}-t} {T_{G}{\tau}_{sf} }}\right)  
\end{equation}

\noindent where $T_{G}$ is the age of the galaxy (i.e. of the oldest SSP) 
and $\tau_{sf}$  is the duration of the burst in units of the
oldest SSP. A common value of extinction is applied to 
stars of all ages, and a \citealt{2000ApJ...533..682C}) attenuation law has 
been adopted ($R_{V }$=4.05). 

To account for emission above 24~\micron, a  component 
coming from colder, diffuse dust, likely heated by star-formation processes, 
has been included in the fitting procedure. It is represented by 
templates of well-studied starburst galaxies (i.e. Arp 220, M82, M83, NGC 1482, 
NGC 4102, NGC 5253 and NGC 7714) and five additional host-galaxy average 
templates obtained recently by \cite{2011MNRAS.414.1082M} from the 
starburst templates of \cite{2006ApJ...653.1129B}. This set of five 
templates have been included since they properly reproduce the relative PAH 
strengths in the average IRS spectra of our ULIRG sample (see F10). 

Regarding the AGN component, we have used the radiative transfer code of 
\cite{2006MNRAS.366..767F}. This model follows the formalism developed by 
different authors (e.g., \citealt{1992ApJ...401...99P}; 
\citealt{1994MNRAS.268..235G}; \citealt{1995MNRAS.273..649E}), where the IR 
emission in AGN originates in dusty gas around the SMBH with a ``flared disk'', ``smooth distribution''. Recently, this model has been widely used and found 
to successfully reproduce the photometric data, including the 9.7~\micron\ 
silicate feature in emission observed for type-I AGN 
(e.g., \citealt{2005A&A...436L...5S}). Recent high-resolution, interferometric 
mid-IR observations of nearby AGN (e.g., \citealt{2004Natur.429...47J}) have 
confirmed the presence of a geometrically thick, torus-like dust 
distribution on pc-scales; this torus is likely irregular or
``clumpy''.
%
Over the last decade, many codes have been developed to deal 
with clumpy dust distributions (e.g., \citealt{2002ApJ...570L...9N}; 
\citealt{2008ApJ...685..160N}; \citealt{2010A&A...523A..27H}). 

According to \cite{2005A&A...436...47D}, the two models (smooth and clumpy) 
do not differ significantly in reproducing sparse photometric 
datapoints (see also \citealt{2008NewAR..52..274E}). The main difference is in 
the strength of the silicate feature observed in absorption in objects 
seen edge-on, which is, on average, weaker for clumpy models with the same 
global torus parameters. 
The comparison between the two models has been applied only to few 
samples including IRS data and the results are not conclusive. 
Recently, \cite{2011MNRAS.416.2068V} made a comparison of the smooth vs. 
clumpy models for the matter responsible for reprocessing the nuclear 
component of a hyper-luminous absorbed \xray\ quasar at 
$z{\sim}0.442$ (IRAS~09104$+$4109), for which both photometric and 
spectroscopic (IRS) data were available. While smooth solutions (the F06 model) 
are able to reproduce the complete dataset, clumpy models 
(\citealt{2008ApJ...685..160N}) have problems in reproducing the source 
photometry and spectroscopy at the same time. 
In \cite{2011MNRAS.414.1082M}, smooth vs. clumpy models are tested 
against a sample of ${\sim}$10 local \swift/BAT AGN with prominent 
emission at \iras\ wavelengths. 
In this case, the authors claim that clumpy solutions reproduce the data 
better; smooth model parameters produce a much wider range of SED solutions, 
i.e. this model seems to produce too degenerate solutions. 
A further, more complete and extensive comparison of smooth vs. clumpy 
solutions is presented in Feltre et al. (submitted), where the theoretical SED 
shapes and the detailed spectral features of the two classes of models 
(i.e. F06 for the ``smooth distribution'' and \citealt{2008ApJ...685..160N} 
for the ``clumpy distribution'') are compared using a large compilation of AGN 
with \irs\ spectroscopic data. Overall, similar results to 
those obtained by \cite{2005A&A...436...47D} are derived, 
i.e. SED fitting applied to both photometric and spectroscopic data is 
not a sufficiently reliable tool to discriminate between the smooth and the
clumpy distributions. 
We remind the reader that in the present paper we are focusing 
on the torus ``global'' energy output (i.e., the relevance of accretion-related 
emission with respect to the total source SED), not on the details of the 
torus structure and geometry, so the choice of the adopted model does not 
critically influence the results; as shown by Feltre et al. (submitted), the two models provide consistent results in terms of 
energetics. 
%
\begin{figure*}
\centering
\includegraphics[width=0.9\textwidth]{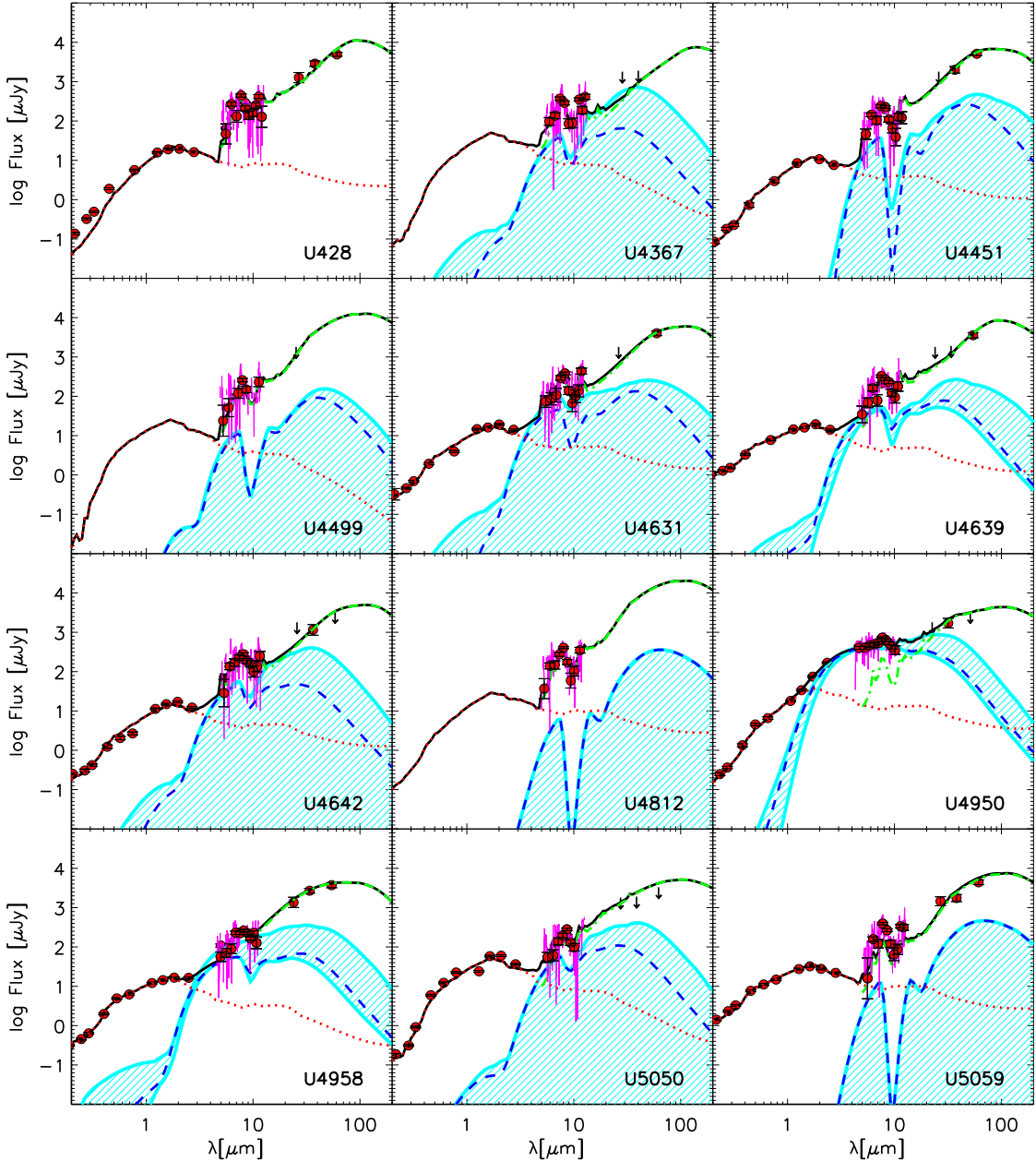}
\caption{{\bf a)} Rest-frame broad-band datapoints (red dots) compared with 
the best-fit model obtained as the sum (solid black line) of a stellar (red 
dotted line), an AGN (blue dashed line) and a starburst component 
(green dot-dashed line). IRS spectra are shown as magenta lines. 
The area filled with diagonal lines represents the AGN solutions at the 
3$\sigma$ confidence level.} 
\label{figure_sed}
\end{figure*}
%
\addtocounter{figure}{-1}
\begin{figure*}
\centering
\includegraphics[width=0.9\textwidth]{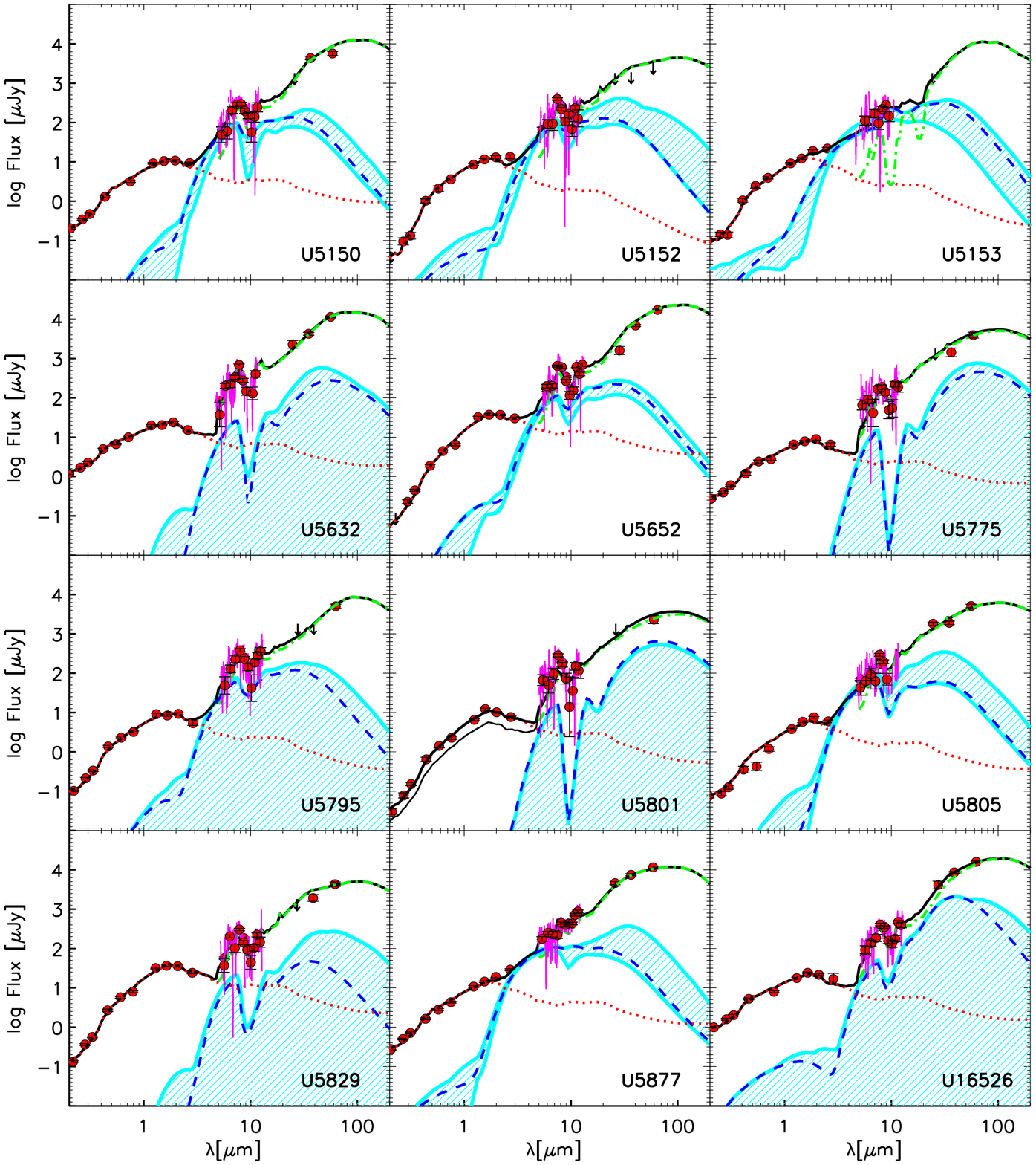}
\caption{{\bf b)} As in Fig.~\ref{figure_sed}a.}
\end{figure*}

\begin{figure*}
\centering
\includegraphics[angle=90,width=1.1\textwidth]{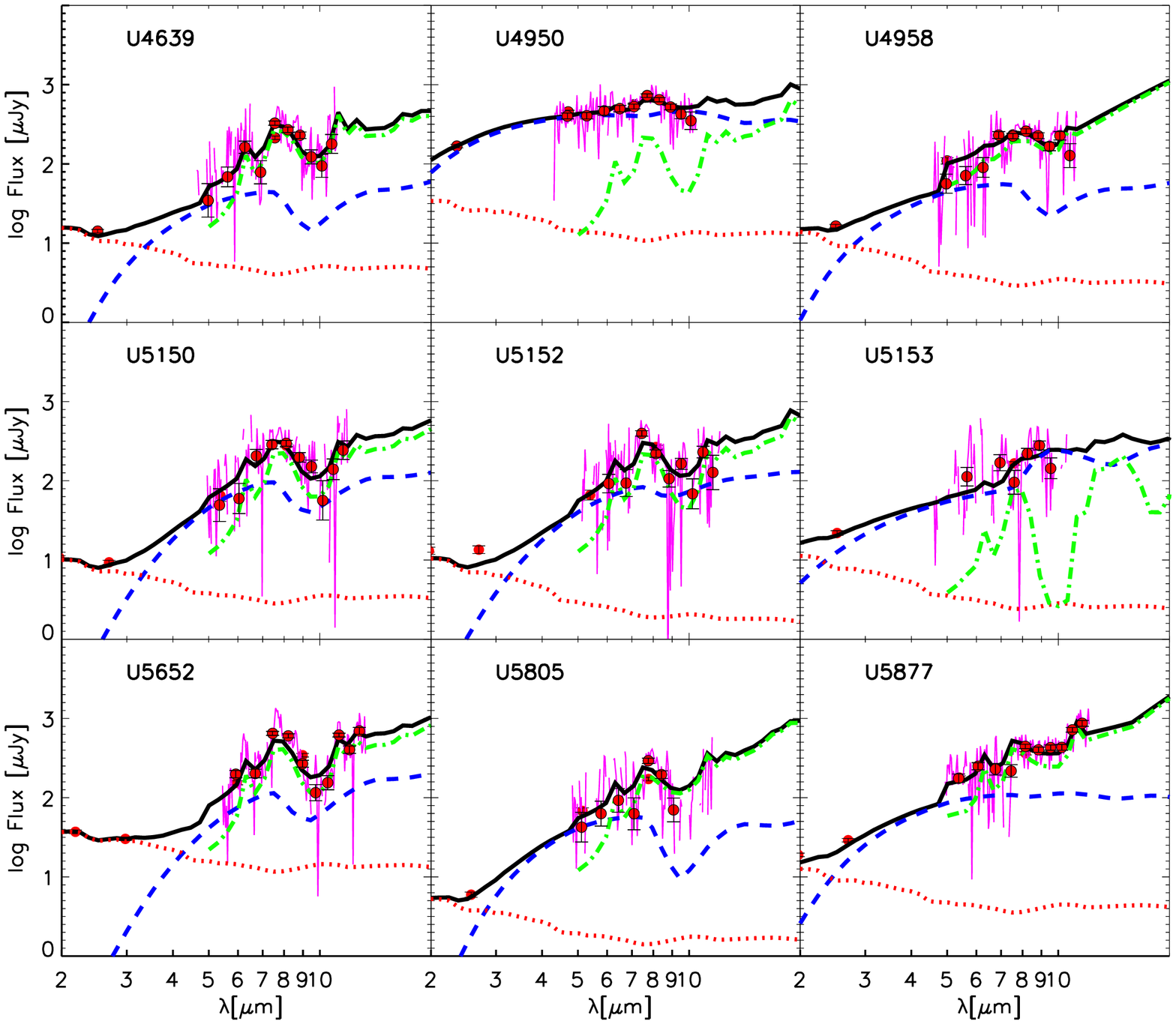}
\caption{Rest-frame SEDs and datapoints, as in Fig.~\ref{figure_sed}, 
where a zoom in the 2--20~\micron\ wavelength range is shown for the 
nine sources where an AGN component is required at the 3 sigma confidence 
level from the SED-fitting analysis.}
\label{figure_sed_2_20_um}
\end{figure*}
%

\subsection{SED-fitting procedure}
\label{fitting_sec}
In most previous work using the F06 code, only photometric datapoints were 
used, and the quality of the fitting solutions was estimated using a standard 
$\chi^{2}$ minimization technique, where the observed values are the 
photometric flux densities (from optical-to-mid-IR/far-IR) and the model values 
are the ``synthetic'' flux densities obtained by convolving the sum of stars, 
AGN, and starburst components through the filter response curves 
(see \citealt{2008MNRAS.386.1252H}). 
In \cite{2011MNRAS.416.2068V}, the spectroscopic information was taken 
into account {\it a posteriori} in order to discriminate among different 
photometric best-fitting solutions.

Here we propose a first attempt to simultaneously fit the photometric and 
spectroscopic datapoints using smooth torus models by transforming the mid-IR 
14--35~\micron\ observed-frame spectra into ``narrow-band'' 
photometric points of 
1~\micron\ band-width in the observed frame (i.e., subdividing the 
spectral transmission curve in sub-units), and estimating the 
corresponding fluxes and uncertainties using ordinary 
procedure of filter convolution and error propagation. 
The ``new'' filters have been chosen to achieve, in each wavelength bin, 
a sufficient signal-to-noise ratio without losing too much in terms of 
spectral resolution, which is needed to reproduce the 9.7~${\mu}$m feature, 
when present. To take into account slit loss effects, 
IRS data have been normalized to the 24~\micron\ flux density, 
deriving normalization factors between $\sim$1.2 and 1.8. 
We would like to remark here that the F06 code does not necessarily impose 
the presence of an AGN component, i.e., solutions with only stellar emission 
are possible, as described in $\S$\ref{agnfraction_sec_fromir}. 
Furthermore, the relative normalization of the optical/near-IR component and 
the far-IR emission of the host galaxy is free, given the extremely complex 
physical relation of the two (i.e. \citealt{2010A&A...518L..39B}).
Overall, the SED-fitting procedure ends with 11 free parameters: six 
are related to the AGN, two to the stellar component, and one to the starburst. 
The further two free parameters are the normalizations of the stellar and 
of the starburst components; the torus normalization is estimated by 
difference, i.e., it represents the scaling factor of the torus model 
capable to account for the data-to-model residuals once the stellar 
components have already been included. 
Here we briefly recall the parameters involved in our SED fitting analysis, 
and refer to F06 and Feltre et al. (submitted) for a detailed description of 
the AGN model parameter.
The six parameters related to the torus are: the ratio $R_{max}/R_{min}$ between 
the outer and the inner radius of the torus (the latter being defined 
by the sublimation temperature of the dust grains); the torus
opening angle $\Theta$; the optical depth $\tau$ at 9.7~\micron\ 
($\tau_{9.7{\mu}m}$); the line of sight $\theta$ with respect to the 
equatorial plane; two further parameters, $\gamma$ and $\alpha$, 
describe the law for the spatial distribution of the dust. 
In the currently adopted version (see Feltre et al., submitted), the ``smooth'' 
torus database contains 2368 AGN models. In 
\cite{2010A&A...517A..11P} (see also \cite{2008MNRAS.386.1252H}), 
the degeneracies related to the six torus parameters are extensively 
described: in fact, various combinations of parameter values are equally 
able to provide good results in reproducing a set of observed data points. 
In particular, the optical depth $\tau_{9.7{\mu}m}$ has the 
largest effect on the fit. The infrared luminosity,
provided by the SED-fitting code, is robustly
determined (within $\sim$0.1 dex) and appears ``solid'' 
against parameter degeneracies.

Concerning the parameters associated with the other components, 
two are related to 
the stellar emission: $\tau_{sf}$, i.e. the parameter of the Schmidt-like law 
for the star formation, and the reddening $E(B-V)$. 
Regarding the starburst 
component, the free parameter is related to the choice of the best-fitting 
template among the starburst library. 

The given number of free parameters means that the acceptable solutions, within 
1(3)$\sigma$ confidence levels, are derived, for each source, by considering 
the parameter regions encompassing $\chi^{2}_{min}$+(12.65, 28.5), respectively, 
in presence of 11 degrees of freedom (d.o.f.; 
see \citealt{1976ApJ...208..177L}). 

\begin{figure}
\includegraphics[width=8cm]{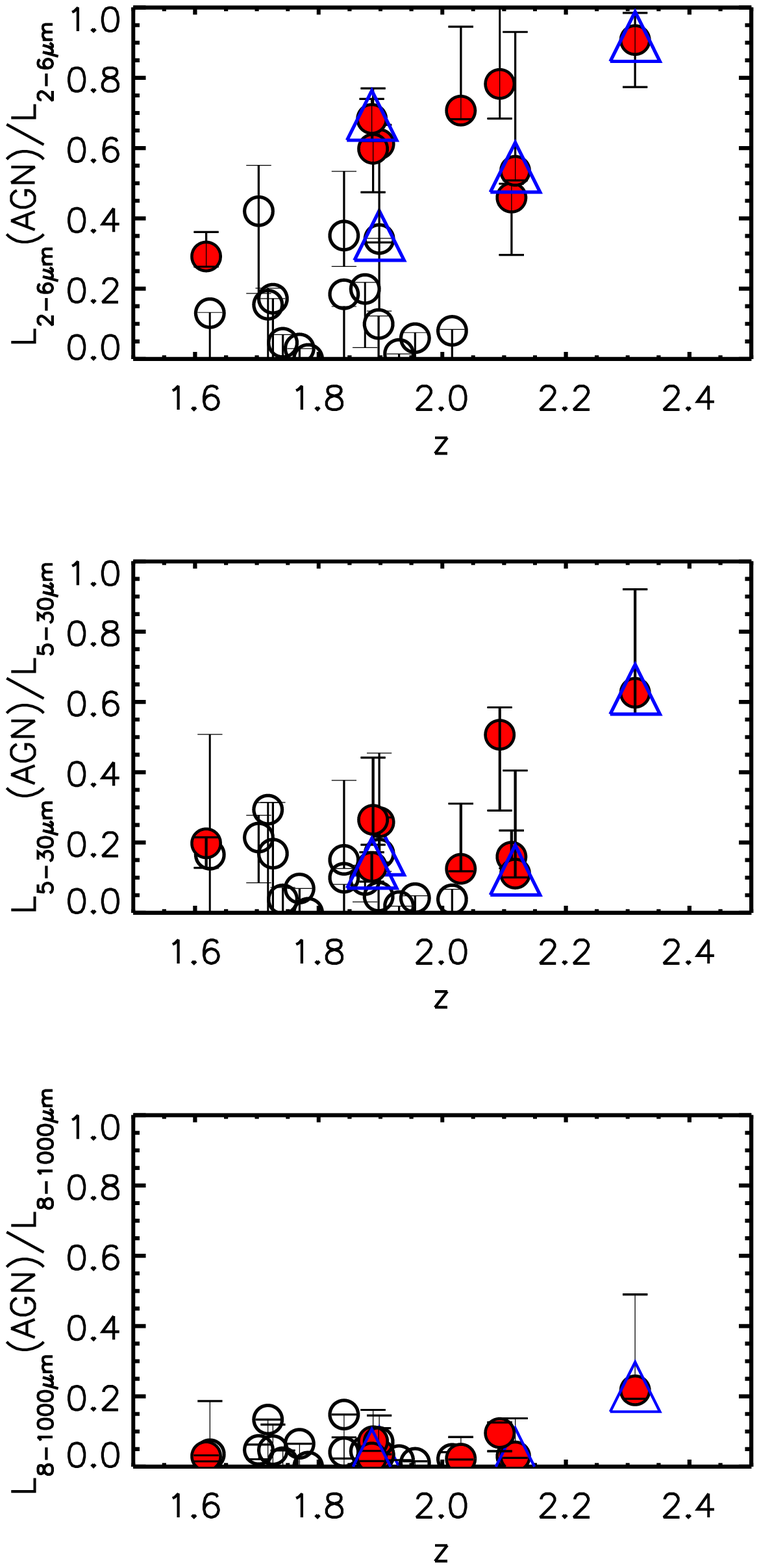}
\caption{Fractional contribution due to the AGN component in the 
2-6~\micron\ ({\it top panel}), 5--30~\micron\ ({\it middle panel}), and
8--1000~\micron\ ({\it bottom panel}) range. The error bars 
account for the AGN model dispersion at the 1$\sigma$ confidence level. 
Red points indicate the sources where an AGN component is detected at 
the 3$\sigma$ confidence level from the SED fitting, and the triangles 
those with \xray\ emission pointing clearly towards an AGN classification.}
\label{figure_agn_contr}
\end{figure}

\section{AGN fraction}
\label{agnfraction_sec}

\subsection{SED-fitting results}
\label{agnfraction_sec_fromir}
In Fig.1a,b the observed UV--160~\micron\ datapoints (filled red points) 
are reported along with the best-fitting solutions (black lines) and the 
range of AGN models within the 3$\sigma$ confidence level (filled region). 
All the sources need a host galaxy (red dotted-line) and a starburst 
component (green dot-dashed line). 
The host galaxy dominates the UV--8~\micron\ photometry (at $z\sim2$, the 
IRAC 8~\micron\ band samples the 2.7~\micron\ rest-frame), while the 
starburst component dominates at longer wavelengths. For the three sources 
with no PACS detection (U5050, U5152, and U5153), the starburst component 
is required by the SED-fitting procedure to reproduce the mid-IR spectral 
data, although its shape is not well constrained. 

Regarding the AGN component, our goal is to check whether its presence 
is required by the data and, in this case, to estimate its contribution to the 
IR luminosity. The SED-fitting procedure found that for all but one
source (U428) the presence of an AGN is consistent with the
photometry, although for only nine sources 
(${\sim}35\%$ of the sample: U4639, U4950, U4958, U5150, U5152, U5153, 
U5652, U5805, and U5877) its presence is significant 
at the 3$\sigma$ confidence level (i.e., solutions with no torus emission 
have ${\chi}^{2}{\ge}{\chi}^{2}_{min}$+28.5). 
However, in these sources as well, the AGN component far from dominates 
the whole spectral range, but emerges only in the narrow 2--10~\micron\ range, 
where the stellar emission from the host galaxy has a minimum while 
the warm dust heated by the AGN manifests itself 
(e.g., \citealt{2000A&A...359..887L}; \citealt{2010A&A...518L..27G}). 

In Fig.~\ref{figure_sed_2_20_um} we report a zoom of the SED over the 
2--20~$\mu$m range in order to visualize the AGN emission for all of the nine 
sources where such component is required at the 3$\sigma$ confidence level. 
At a visual inspection, the presence of a nuclear component can be inferred 
for sources with a power-law SED (i.e., U4950 and U5877), for sources where 
the stellar component alone is not able to reproduce the datapoints around 
3~$\mu$m (i.e., U4958, U5153) or for sources where the starburst component, 
normalized to the far-IR datapoints, has already declined around 5--6~$\mu$m 
(i.e. U5152, U5153, and U5805). Finally, there are few cases (i.e., U4639) 
where visual inspection of the SED decomposition is not strongly suggesting 
an AGN component, which is however required by the SED-fitting analysis.

The likely AGN in these sources is either obscured or of low-luminosity, 
although a combination of both effects is plausible as well. 
While our analysis is able to place constraints on the presence of an AGN, 
we cannot draw any firm conclusion on either the obscured or the low-luminosity 
hypothesis for the AGN emission from mid-IR data. 
Despite the uncertainties affecting the estimate of the gas column density 
$N_{H}$ derived from the dust optical depths (e.g., 
\citealt{2001A&A...365...28M}), what we observe is that for all of the 
sources with an AGN component, a certain level of obscuration\footnote{The 
column density has been derived from the the optical depth at 9.7\micron, 
using the Galactic extinction law (\citealt{1989ApJ...345..245C}) 
and dust-to-gas ratio (\citealt{1978ApJ...224..132B}).} 
($10^{22}$\simlt$N_{H}$\simlt$10^{24}$~cm$^{-2}$) is required. 

In particular, the three sources where the presence of an AGN is particularly 
evident in the mid-IR regime from our SED decomposition
(U4950, U4958, and U5877) -- as already pointed out by F10 
on the basis of the presence of a powerlaw mid-IR SED (U4950 and U5877), 
optical emission lines (U4958, with an apparently 
broad C\ {\sc iii}] feature) and optically unresolved nucleus 
(U4950 and U5877) -- also show a relatively bright \xray\ counterpart (see 
Sec.~\ref{xray_section}). The optical depth at 9.7~\micron\ of the
best-fitting solutions corresponds to $N_{H}\sim{(2-7)}{\times}10^{22}$~cm$^{-2}$
(i.e., in the Compton-thin regime; $<N_{H}>=6{\times}10^{22}$~cm$^{-2}$ is 
obtained once the solutions at the 3$\sigma$ level are considered). 
Of the remaining six sources, one (U4639) has an association with a 
relatively weak \xray\ source, while for the others only an upper limit to the 
\xray\ emission can be placed (see Table~\ref{xray_properties}). 
For these six sources, the SED-fitting procedure indicates an obscuration 
still in the Compton-thin regime, but higher 
(up to $4{\times}10^{23}$~cm$^{-2}$, with $<N_{H}>=2{\times}10^{23}$~cm$^{-2}$ 
when the solutions at the 3$\sigma$ 
level are taken into account) than for the previous three sources. 

Further insights on the properties of these sources, hence on the nature of 
their broad-band emission, will be discussed using \xray\ diagnostics 
(see Sec.~\ref{xray_section}). 
 
Turning now to source energetics, for the nine objects with an AGN component 
from the mid-IR, we computed the total and nuclear luminosity in three 
different spectral ranges: 
the whole (8--1000~$\mu$m) IR range, the mid-IR (5--30~$\mu$m) 
range (partially sampled by IRS), and the narrow 2--6~$\mu$m 
wavelength interval. 
We find that the 8--1000~$\mu$m luminosity is always completely 
dominated by star-formation emission, the AGN nuclear contribution being 
$\simlt$5$\%$ (see Fig.~\ref{figure_agn_contr}, {\it bottom panel}), 
and that in only one source (U4950) out of the nine is the nuclear component 
contribution significant (${\sim}20\%$). 
Our finding that starburst processes dominate the 8--1000 $\mu$m 
emission is consistent with the F10 and N12 conclusions. 

In the 5--30~\micron\ range, where the re-processed emission from the dust 
surrounding the nuclear source peaks (e.g., \citealt{2004MNRAS.355..973S}), 
we find a larger AGN contribution (i.e. ${\sim}$25\%; see 
Fig.~\ref{figure_agn_contr}, {\it middle panel}). As shown in Fig.1a,b, 
at ${\lambda}{\sim}10$~${\mu}$m the nuclear component typically 
starts being overwhelmed by the starburst emission. 
We note that the mid-IR AGN/starburst relative contribution is also discussed 
by F10, adopting a completely different method than ours (i.e., scaling the SED 
of Mrk~231 to the 5.8~$\mu$m continuum and fitting the residual SED with the 
average starburst from \citealt{2006ApJ...653.1129B}). 
Excluding from their analysis the three sources with the strongest 
AGN evidence from either \xray\ or optical data (U4950, U4958, and U5877), 
F10 found a nuclear fraction of $\sim$20\% (see their Fig.~22, 
{\it top panel}), which is consistent with our results. 

The only wavelength range where we find that the AGN overcomes the galaxy 
emission and contributes to ${\sim}$60\% of the emission is the 
narrow 2--6~${\mu}$m interval (see Fig.~\ref{figure_agn_contr}, 
{\it top panel}). The power of the near-IR spectral range to detect 
obscured AGN emission confirms previous results (e.g., 
\citealt[2010]{2006MNRAS.365..303R} using $L$-band spectroscopy, 
but see also \cite{2003ApJ...588..199L}, where a weak near-IR excess continuum 
emission, detected in disk galaxies thanks to ISOPHOT spectral observations, 
was ascribed to interstellar dust emission at temperatures of ${\sim}10^{3}$K). 

The dominance of the AGN emission in the narrow 2--6~${\mu}$m 
interval is not in contrast with N12 conclusions, i.e. that these 
sources in the mid-IR are dominated by the PAH features, once the different 
spectral range is taken into account. 

In Table~\ref{irx_lum}, we report the total IR luminosities 
(8--1000~\micron), the AGN fractions over the three wavelengths discussed 
above (for the nine sources with an AGN emission detected at least at the 
3$\sigma$ confidence level), and both observed and predicted 2--10~keV
luminosities for our sample. 
The latter luminosities have been derived from the 5.8~\micron\ 
luminosity, using the \cite{2009ApJ...693..447F} correlation (see
Sec.~\ref{xray_section} for further details), and only for the
nine sources with an AGN component detected at least at
3$\sigma$ confidence (last column of Table~\ref{irx_lum}). 
The integration of the SED over the 8--1000~\micron\ range confirms the 
ULIRG classification ($L_{IR}>10^{12}L_{\odot}$), inferred by F10 on the basis 
of the 24~\micron\ flux densities and redshifts, for 14 out of the 24 sources, 
the remaining 10 sources showing slightly lower IR luminosities, 
between 0.5${\times}10^{12}$ and $10^{12}~L_{\odot}$. The fact that 
24~${\mu}$m--based measurements tend to over-estimate the $L_{IR}$ for 
$z{\sim}2$ sources is consistent with other works based on {\it Herschel}-PEP 
data (e.g., \citealt{2010A&A...518L..29E}; N12) and stacking methods 
(e.g., \citealt{2007ApJ...668...45P}). 
The final IR luminosity range of our sample is 
0.5--2.8${\times}10^{12}~L_{\odot}$ ($<L_{IR}>=1.4{\times}10^{12}~L_{\odot}$, 
with a dispersion $\sigma=7{\times}10^{11}~L_{\odot}$).

\subsection{X-ray results}
\label{xray_section}

\begin{table*}
\begin{center}
\caption{X-ray properties of the sample of $z\sim2$ IR-luminous galaxies}
\label{xray_properties}
\footnotesize
\begin{tabular}{lcccccccc}
\hline\hline
Name & XID & F$_{0.5-8~keV}$ & log(L$_{2-10~keV}$)
&log(L$_{2-10~keV}$,fit)  &  $N_{H}$ & Class. X (Xu et al. 11)&Class. X (revised) & Class. SED \\
(1)            & (2) &  (3)  &  (4)  &  (5)  & (6) &(7)&(8)&(9)\\
\hline
\multicolumn{5}{c}{\sf Sources with an \xray\ counterpart in the 4~Ms CDF-S catalog} \\ \\
U428           & 579 & 0.07  &  41.9 & -        & -   &                                                 Gal & Gal &Gal.\\
U4639          & 555 & 0.07  &  42.0 & -     & - &                                                  Gal & Gal &AGN\\                           
U4642          & 437 & 0.29  &  42.5 & 42.8 &4.5$^{+6.1}_{-4.3}\times{10^{22}}$   & AGN&AGN&Gal.\\
U4950          & 351 & 6.64  &  43.9 & 44.2  &  1.2$^{+0.3}_{-0.2}\times{10^{23}}$&AGN&AGN&AGN$^{\star}$\\ 
U4958          & 320 & 0.31  &  42.3 & 43.7  & 1.9$^{+1.4}_{-0.7}\times{10^{24}}$& AGN&AGN&AGN$^{\star}$\\ 
U5632          & 552 & 0.09  &  42.1 & -     & -   &                                                    AGN&Gal.&Gal.\\
U5775          & 360 & 0.05  &  41.8 & -     & -  &                                                     AGN&Gal.&Gal.\\
U5877         & 278 & 3.34  &  43.0 & 44.0  &6.0$^{+2.0}_{-1.5}\times{10^{23}}$ &AGN&AGN&AGN$^{\star}$\\ 
\hline
\multicolumn{5}{c}{\sf Sources without \xray\ detection in the 4~Ms catalog} \\ \\
U4367  & - & - & $<42.3$ & - & - & - & - &Gal.\\
U4451  & - & - & $<42.2$ & - & - & - & - &Gal.\\
U4499  & - & - & $<42.1$ & - & - & - & - &Gal.\\
U4631  & - & - & $<42.0$ & - & - & - & - &Gal.\\
U4812  & - & - & $<42.4$ & - & - & - & - &Gal.\\
U5050  & - & - & $<42.2$ & - & - & - & - &Gal.\\
U5059  & - & - & $<42.0$ & - & - & - & - &Gal.\\
U5150 & - & - & $<42.4$ & - & - & - & - &AGN\\
U5152  & - & - & $<42.5$ & - & - & - & -&AGN\\
U5153  & - & - & $<42.6$ & - & -& - & - &AGN\\
U5652 & - & - & $<41.8$ & - & -& - & - &AGN\\
U5795  & - & - & $<41.8$ & - & -& - & -  &Gal.\\
U5801  & - & - & $<41.9$ & - & - & - & - &Gal.\\
U5805  & - & - & $<42.1$ & - & - & - & - &AGN\\
U5829  & - & - & $<41.9$ & - & -& - & -  &Gal.\\
U16526 & - & - & $<42.2$ & - & - & - & - &Gal.\\
\hline
\hline   
\end{tabular}
\begin{minipage}[l]{0.94\textwidth}
\footnotesize
Notes ---  
(1) Source name (F10); 
(2) XID from \cite{2011ApJS..195...10X}; 
(3) observed-frame 0.5--8~keV flux in units of 10$^{-15}$~erg~cm$^{-2}$~s$^{-1}$ 
(\citealt{2011ApJS..195...10X}); errors on the \xray\ fluxes (hence on 
the luminosities) vary from less than 10\% for the few sources with most 
counts up to $\sim$~40\% for the \xray\ faintest sources; 
(4) logarithm of the rest-frame 2--10~keV luminosity derived from either the 
0.5--8~keV flux or the 0.5--8~keV sensitivity map (upper limits in the 
lower panel). The flux (or upper limit) is converted into a luminosity 
by assuming a powerlaw model with $\Gamma$=1.4;  
(5) logarithm of the rest-frame, absorption-corrected 2--10~keV luminosity, 
in units of erg~s$^{-1}$; (6) column density. 
Both (5) and (6) have been derived directly from \xray\ spectral fitting, 
which has been limited to the four sources with the highest counting 
statistics (see Sect.~\ref{xray_section} for details). 
We note that \xray\ luminosities -- columns (4) and (5) -- are 
different because of the different spectral modeling and assumptions 
(see \citealt{2011ApJS..195...10X} and Sect.~\ref{xray_section}); 
(7), (8), (9) source classifications from \xray\ data 
(\citealt{2011ApJS..195...10X} and current work) and from SED-fitting, 
respectively. The term ``AGN'' in the source classification based 
on SED fitting (last column) indicates that the nuclear component is detected 
at the 3$\sigma$ confidence level. $^\star$ 
indicates the presence of an AGN from optical data (U4958 -- optical
emission lines; U4950 and U5877 -- optical morphology, see F10).
\end{minipage}
\end{center}
\end{table*}

The 4~Ms \xray\ source catalog in the CDF-S (\citealt{2011ApJS..195...10X}) 
provides additional information for the sub-sample of eight matched sources 
(see Table~\ref{xray_properties}). 
The depth of the \chandra\ mosaic in the field, though variable across its area,
provides constraints down to very faint flux limits 
($\sim10^{-17}$~\cgs\ in the 0.5--2~keV band). 
In the following, we use the \xray\ information to provide an independent 
estimate of the AGN presence in our sample of IR-luminous galaxies and, 
for sources with most counts, characterize their \xray\ emission. 
We also provide upper limits to the \xray\ luminosity of the sources 
which are not detected in the 4~Ms CDF-S image. 

A basic source classification is reported by \cite{2011ApJS..195...10X} 
(AGN/galaxy/star), where five different classification criteria are adopted 
to separate AGN from galaxies\footnote{
The criteria to classify a source as an AGN are: 
high \xray\ luminosity (above $3\times10^{42}$~\lum); 
flat effective photon index ($\Gamma\leq1.0$); 
\xray-to-optical flux ratio log(f$_{\rm X}$/f$_{\rm R}$)$>-1$; 
an \xray\ luminosity at least three times higher than that possibly due to star 
formation (see \citealt{2005ApJ...632..736A}); 
presence of either broad or high-ionisation emission lines in the 
optical spectra. See $\S$4.4 of \citealt{2011ApJS..195...10X} for details.}; 
at least one of these criteria should be satisfied. According to such 
classification, only two sources of the present sample are classified as 
galaxies, U428 and U4639, the remaining six being AGN. 
However, in the following we will provide indications for a 
``revision'' of this classification using all of the available \xray\ 
information (e.g., spectral properties, luminosities, count
distribution). 

 In Table~\ref{xray_properties} the source classifications from \xray\
data (\citealt{2011ApJS..195...10X} and current work, respectively), along with 
the results from the SED-fitting analysis (see Sec.~\ref{agnfraction_sec_fromir}), are presented.

X-ray luminosities, coupled with \xray\ spectral analysis (see below), 
indicate the presence of an AGN in the three sources (U4950, U4958, and 
U5877) for which the SED fitting analysis and F10 already suggested an AGN. 
Also U4642 has an \xray\ luminosity typical of AGN emission.

In comparison with the 2~Ms \chandra\ data (\citealt{2008ApJS..179...19L}) 
used by F10, U4958 represents a new AGN detection, although the presence of an 
active nucleus was already inferred by the mid-IR featureless spectra and the 
optical spectrum, showing a broad \mbox{C\ {\sc iii}]} line and strong 
\mbox{N\ {\sc v}} and \mbox{C\ {\sc iv}} emission lines.

All of the \xray\ matched sources have a full-band (0.5--8~keV) detection. 
One source (U4958) has an upper limit in the soft band (0.5--2~keV); 
this result is suggestive of heavy absorption, since \chandra\ has its highest
effective area, and hence best sensitivity, in this energy range. 
Basic \xray\ spectral analysis (using an absorbed powerlaw with photon index 
fixed to 1.8, as typically observed in AGN; \citealt{2005A&A...432...15P}), 
actually confirms the presence of strong, possibly Compton-thick 
obscuration\footnote{A source is called Compton-thick if its column density 
$N_{\rm H}>1/\sigma_{\rm T}\sim1.5\times10^{24}$~cm$^{-2}$, where $\sigma_{\rm T}$ 
is the Thomson cross section.} towards this source 
(N$_{\rm H}=1.9^{+1.4}_{-0.7}\times10^{24}$~cm$^{-2}$). 
Four sources have hard-band (2--8~keV) upper limits: U428, U4639, U5632, 
and U5775, all of which being characterized by limited counting statistics 
(12--26 counts in the full band). Their \xray\ luminosity 
(see Table~\ref{xray_properties})\footnote{For the four sources with 
2--8~keV upper limits, the 2--10~keV luminosities are reported in 
Table~\ref{xray_properties} as these sources were detected in the hard band. 
We note, in fact, that the values reported in the table are derived from 
the 0.5--8~keV fluxes, where all of these sources are detected.} 
is consistent with star-formation activity, although the presence of 
a low-luminosity AGN cannot be excluded.

U4950 and U5877 are characterized by $\sim$1420 and $\sim$360 counts 
in the 0.5--8~keV band, which allow a moderate-quality \xray\ spectral 
analysis. Both sources can be fitted using a powerlaw model, 
but the flat photon index ($-1$ and $+$1, respectively) is indicative of 
absorption, which has been estimated to be 
1.2$^{+0.3}_{-0.2}\times10^{23}$~cm$^{-2}$ and 
6.0$^{+2.0}_{-1.5}\times10^{23}$~cm$^{-2}$, respectively, after fixing the photon 
index to 1.8. 
The same model applied to U4642 data provides good results, although 
the derived column density is poorly constrained ($N_{\rm H}=4.5^{+6.1}_{-4.3}\times10^{22}$~cm$^{-2}$). Apart from U4950, U4958, U5877 and U4642, for the remaining \xray\ matched sources, 
the limited counting statistics prevent us from placing constraints on any 
possible obscuration. 

For the sources which were not detected in the 4~Ms CDF-S observations, 
we derived rest-frame 2--10~keV luminosity upper limits by converting 
the full-band sensitivity map (weighted over a region surrounding the source 
of interest of 4\arcsec\ radius to minimize possible spurious fluctuations in 
the map) using a powerlaw with $\Gamma$=1.4 (see Table~\ref{xray_properties}). 
The assumption of $\Gamma$=1.8 would imply luminosities higher 
by $\sim0.1-0.2$ dex. 
The derived \xray\ upper limits place strong constraints on the further 
presence of luminous AGN in our sample, unless they are heavily
obscured.  

Following F10, for the nine sources with an AGN from the SED fitting, 
we estimated the rest-frame \xray\ emission using the mid-IR emission
of the AGN component as a proxy of the nuclear power 
(Table~\ref{irx_lum}). In particular,  we use the $L_{2-10keV}$--5.8~\micron\ relation 
(\citealt{2009ApJ...693..447F}; 
$\log L_{2-10keV}=43.57 + 0.72\times (\log L_{5.8\micron}-44.2)$, 
with luminosities expressed in erg~s$^{-1}$; see also 
\citealt{2004A&A...418..465L}), which was calibrated using 
unobscured AGN in the CDF-S and COSMOS surveys.

\begin{figure}
\includegraphics[width=8cm]{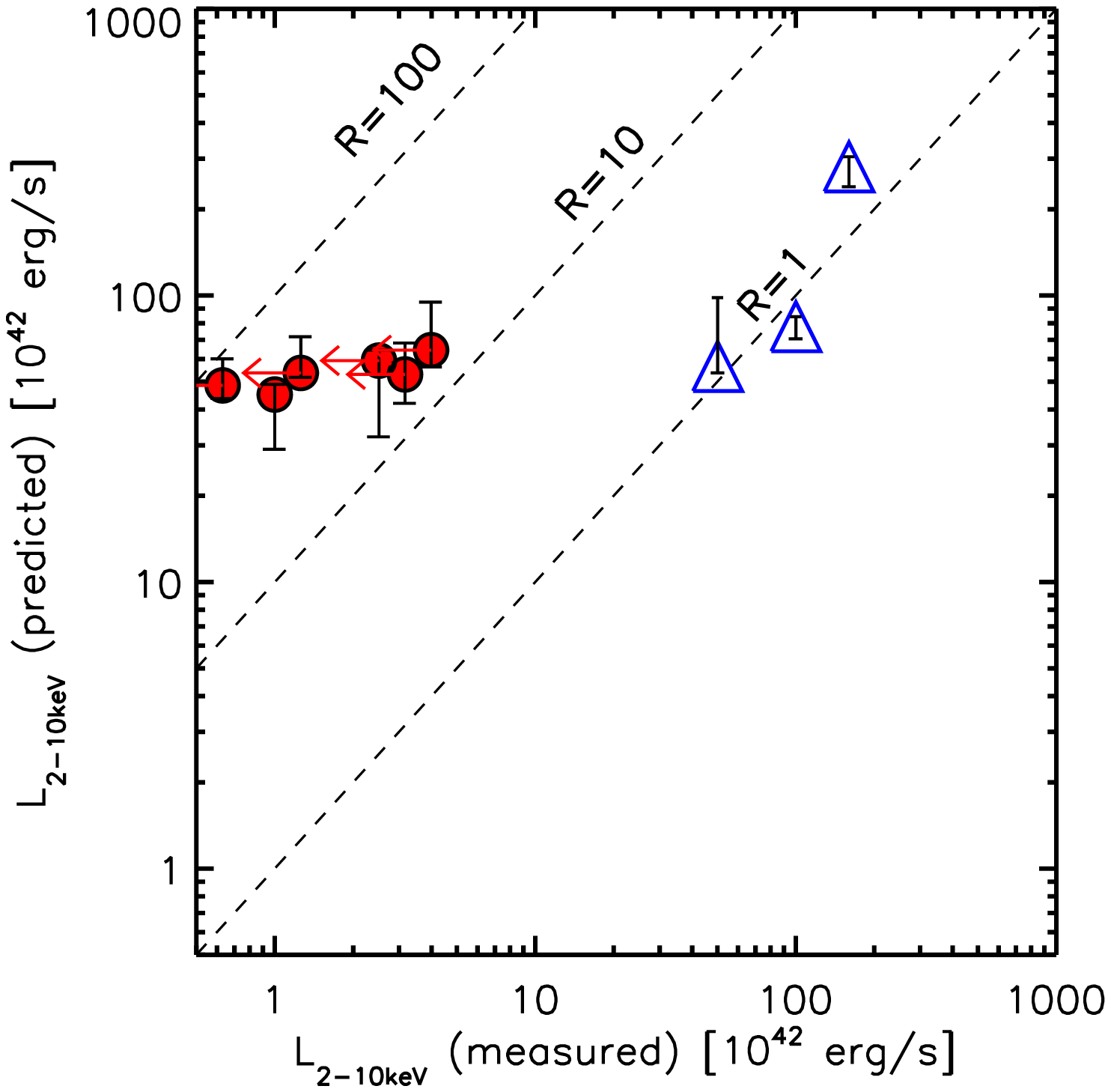}
\caption{Predicted vs. measured 2--10~keV luminosities for the nine 
sources where an AGN component is inferred from the SED fitting. The 
triangles indicate the three sources with clear AGN \xray\ emission; 
for these three sources only, the 2--10~keV luminosities have been 
corrected for absorption derived from \xray\ spectral analysis 
data (see Table~\ref{xray_properties}). 
The dashed diagonal lines indicate ratios of 1:1, 10:1 and 100:1 between the 
predicted and the measured \xray\ luminosity (from right to left).}
\label{lxlx}
\end{figure}

In Fig.~\ref{lxlx} the predicted and measured \xray\ 
luminosities are compared. For the three sources with clear AGN signatures 
(also from \xray\ data), the luminosities have been corrected for the 
absorption (see Table~\ref{xray_properties} and above), while for the remaining 
sources, no correction has been applied, due to the lack of constraints to 
any possible column density from the low-count \xray\ spectra.
Overall, it appears that the three sources showing 
clear AGN signatures in the mid-IR band and in X-rays (U4950, U4958 and U5877) 
are broadly consistent with the 1:1 relation, considering the dispersion 
in the mid-IR/\xray\ relation, the observational errors, and 
the uncertainties in deriving the intrinsic 2--10~keV luminosity properly. The correction accounting for the absorption seems to be correct also 
for the most heavily obscured of our \xray\ sources, U4958.

For the other six sources, the expected 2--10~keV 
luminosity is a factor of ${\sim}$10--100 higher than the measured \xray\ 
luminosity. 
Although caution is obviously needed in all the cases where mid-IR vs. 
\xray\ luminosity correlations are used (see, e.g., the discussion in 
\citealt{2010MNRAS.404...48V}), the obscuration derived from 
the SED fitting for the remaining six sources 
($\sim2{\times}$10$^{23}$~cm$^{-2}$) and the ratio between the measured and 
predicted luminosities suggest the presence of significant obscuration
in these sources. Our conclusion is robust against the possible presence of
absorption affecting also the $L_{5.8\micron}$ values, since in this
case the predicted \xray\ luminosities should be considered as lower
limits.
%
Sensitive \xray\ data over a larger bandpass would be needed 
to test the hypothesis of heavy obscuration in these sources. 

\begin{table*}
\begin{center}
\caption{IR and \xray\ luminosities, and AGN fractions for the sample 
of $z\sim2$ IR-luminous galaxies} 
\label{irx_lum}
\footnotesize
\begin{tabular}{lcccccccc}\hline\hline
Name & $z_{IRS}$ & $L_{IR}$ & f$^{AGN}_{8-1000{\mu}m}$ &
f$^{AGN}_{5-30{\mu}m}$&f$^{AGN}_{2-6{\mu}m}$ &log(L$_{2-10keV}$)  &
logL($_{2-10keV}^{pred}$) & Class.SED \\
(1) & (2) & (3) & (4) & (5) & (6) &(7) &(8)&(9)\\
\hline
   U428  &    1.78   &    12.06   &  - &  -& - &  41.9   &     - &\\
  U4367  &    1.62   &    11.70   &    -        &      -         &     -        &  -  &     - &\\
  U4451  &    1.88   &    12.00   &     -        &    -        &     -        &  -   &     -  &\\
  U4499  &    1.96   &    12.25   &      -        &    -        &     -        &  -   &     - & \\
  U4631  &    1.84   &    11.89   &    -         &    -         &     -        &  -   &     -  &\\
  U4639  &    2.11   &    12.08   &   0.03  &    0.16     &  0.46 &  42.0   &     43.7& AGN \\
  U4642  &    1.90   &    11.83   &   -   &      -  &       -     &  42.8   &     -  &\\
  U4812  &    1.93   &    12.45   &   -     &    -     &   -&  -   &     -  &\\
  U4950 &    2.31   &    12.17   &   0.22   &   0.63   &   0.91   &  44.2   &     44.4 &AGN  \\
  U4958  &    2.12   &    12.10   &   0.03   &   0.11   &   0.54   &  43.7   &     43.7  &AGN \\
  U5050  &    1.73   &    11.90   &    -         &     -     &     -   &  -   &     -  &\\
  U5059  &    1.77   &    11.95   &   -        &   -         &     -       &  -   &  -   &  \\
  U5150  &    1.90   &    12.24   &   0.04   &   0.26   &   0.61   &  -   &     43.8  &AGN  \\
  U5152  &    1.89   &    11.94   &   0.07   &   0.26   &   0.60   &  -   &     43.7 &  AGN \\
  U5153  &    2.09   &    12.26   &   0.10   &   0.51   &   0.78   &  -  &     43.8  & AGN \\
  U5632  &    2.02   &    12.39   &     -        &    -       &   -          &  42.1   &     -  & \\
  U5652  &    1.62   &    12.37   &   0.03   &   0.20   &   0.29   &  -   &     43.7  &AGN  \\
  U5775  &    1.90   &    11.98   &   -       &    -        &   -      &  41.8   &     -  & \\
  U5795  &    1.70   &    11.93   &     -      &     -      &     -      &  -   &     - & \\
  U5801  &    1.84   &    11.70   &      -      &     -       &    -        &  -   &     - &  \\
  U5805 &    2.03   &    12.12   &   0.02   &   0.13   &   0.71   &  -   &     43.7 &  AGN \\
  U5829  &    1.74   &    11.90   &     -      &     -      &     -      &  -   &     - &  \\
  U5877  &    1.89   &    12.37   &   0.03   &   0.13   &   0.68   &  44.0   &     43.9 &AGN   \\
 U16526  &    1.72   &    12.38   &     -     &     -     &      -      &  -   &     -  & \\
 \hline\hline
\end{tabular}
\begin{minipage}[h]{12.2cm}
\footnotesize
Notes ---
(1) Source name (F10);
(2) \irs\ redshift (F10); 
(3) logarithm of the total (AGN$+$stellar) IR (8--1000~\micron) luminosity 
(in $L_{\odot}$); 
(4), (5), (6) AGN contribution to the 8--1000, 5--30 and 2--6~${\mu}$m
luminosity, respectively, for the nine sources with an AGN component at the 
3~$\sigma$ level; (7) logarithm of the 2--10~keV luminosity 
(erg~s$^{-1}$; see Sec.~\ref{xray_section} and Table~\ref{xray_properties}); 
(8) logarithm of the 2--10~keV luminosity predicted from the 
$L_{2-10keV}$--$L_{5.8{\mu}m}$ relation for sources with an AGN
component detected at the 3$\sigma$ confidence level (erg~s$^{-1}$; see 
\citealt{2009ApJ...693..447F} and Sec.~\ref{xray_section});
 (9) Source classification from SED fitting. ``AGN'' indicates that 
the nuclear component is detected at the 3$\sigma$ confidence level. 
All of the remaining sources are consistent with no significant AGN emission 
from the SED-fitting analysis.
\end{minipage}
\end{center}
\end{table*}

\section{Summary}

In this paper we have studied the multi-band properties of a sample of 
IR luminous sources at $z{\sim}2$ in order to estimate the AGN contribution 
to their mid-IR 
and far-IR emission. 
The sample was selected by F10 at faint 24~$\mu$m flux densities 
($S(24{\mu}m){\sim}0.14-0.55$~mJy) and $z=1.75-2.40$ to specifically
target luminosities around $10^{12}~L_{\odot}$, 
i.e. sampling the knee of the IR luminosity function. 

We have extended the previous analysis by taking advantage of new 
far-IR data recently obtained by the \herschel\ satellite as part of 
the guaranteed survey ``PACS Evolutionary Probe'' 
(PEP, \citealt{2011A&A...532A..90L}), and of the recently published 
4~Ms \chandra\ data (\citealt{2011ApJS..195...10X}).
The available photometry, coupled with \irs\ mid-IR data, have been used 
to reconstruct the broad-band SEDs of our IR-luminous galaxies. 
These SEDs have been modeled using a SED-fitting technique with 
three components, namely a stellar, an AGN, and a starburst component. 
The most up-to-date smooth torus model by F06 have been adopted for the 
AGN emission. 
The major results of the work can be summarized as follows.

\begin{itemize}
\item{SED fitting indicates that emission from the host galaxy in the 
optical/near-IR and star formation in the mid-IR/far-IR is required 
for all of the sources. Presence of an AGN component is consistent with 
the data for all but one source (U428), although only for 9 out of the 24 
galaxies (${\sim}35\%$ of the sample) is this emission significant at least at 
the 3$\sigma$ level.}

\item{Of the sub-sample of nine sources that likely harbour 
an AGN according to the SED fitting, we find that their total 
(8--1000~${\mu}$m) and mid-IR (5--30~\micron) emissions are dominated 
by starburst processes, with the AGN-powered emission accounting for only 
$\sim5\%$ and $\sim23\%$ of the energy budget in these wavelength ranges, 
respectively. We find that the AGN radiation overcomes the stellar +
starburst  components only in the narrow 2--6~$\mu$m range, where it 
accounts for $\sim$60\% of the energy budget. In this wavelength range,
stellar emission has significantly 
declined and emission from PAH features and starburst emission is not 
prominent yet.}

\item{For this same sub-sample, the gas column densities (derived by converting 
the dust optical depth at 9.7~$\mu$m obtained from the SED fitting) 
are indicative of a significant level of obscuration. 
In particular, three sources, also detected as relatively bright \xray\ 
sources (U4950, U4958, and U5877, see below), have 
$<N_{H}>=6{\times}10^{22}$~cm$^{-2}$ (considering all the solutions at 
the 3$\sigma$ confidence level), while the remaining six sources 
have $<N_{H}>={2}{\times}10^{23}$~cm$^{-2}$.}

\item{X-ray analysis confirm that three sources (U4950, U4958, and U5877) 
are actually powered by an AGN at short wavelengths, and that this AGN 
varies from being moderately (U4950 and U5877) to heavily obscured, 
possibly Compton thick (U4958). 
The \xray\ luminosity of U4642 is also suggestive of moderately obscured 
AGN emission. 
The remaining four sources detected by \chandra\ have \xray\ emission 
consistent with star-formation processes.}

\item{For the six sources where the AGN is required only by SED 
fitting (i.e., no strong AGN emission is observed in \xray), we 
estimate an intrinsic \xray\ nuclear luminosity from the AGN continuum 
at 5.8 ${\mu}$m. The ratio (from 10 up to 100) between the predicted and the 
measured luminosities suggests the presence of significant obscuration in 
these sources.}

\end{itemize}

\section*{acknowledgements}
The authors thank the referee for his/her useful comments. 
FP and CV thank the Sterrenkundig Observatorium (Universiteit Gent), 
in particular Prof. M. Baes and Dr. J. Fritz, for their kind hospitality. 
The authors are grateful to F.E. Bauer and F. Vito for their help with 
\chandra\ spectra. 
CV acknowledges partial support from the Italian Space Agency (contract 
ASI/INAF/I/009/10/0). \\
PACS has been developed by a consortium of institutes led by MPE (Germany) and 
including UVIE (Austria); KU Leuven, CSL, IMEC (Belgium); CEA, LAM (France); 
MPIA (Germany); INAF- IFSI/OAA/OAP/OAT, LENS, SISSA (Italy); IAC (Spain). 
This development has been supported by the funding agencies BMVIT (Austria), 
ESA-PRODEX), CEA/CNES (France), DLR (Germany), ASI/INAF (Italy), 
and CICYT/MCYT (Spain).

\bibliographystyle{mn2e}
\bibliography{pozzi}

\end{document}